\documentclass[traditabstract]{aa}

\usepackage{graphicx}
\usepackage{txfonts}
\usepackage{longtable}
\usepackage{lscape}
\usepackage{natbib}
\bibpunct{(}{)}{;}{a}{}{,}

\begin{document}
\title{The Hamburg/ESO R-process Enhanced Star survey (HERES) \thanks{Based on observations collected at the European Southern Observatory, Paranal,
    Chile (Proposal numbers 170.D-0010, and 280.D-5011).}}
\subtitle{VI. The Galactic Chemical Evolution of Silicon}

   \author{L. Zhang
          \inst{1,2}
          \and T. Karlsson
          \inst{3}
          \and N. Christlieb
          \inst{4}
          \and A. J. Korn
          \inst{5}
          \and P. S. Barklem
          \inst{5}
          \and G. Zhao
          \inst{1}}

   \institute{Key Laboratory of Optical Astronomy, National Astronomical Observatories, CAS, 20A Datun Road, Chaoyang
              District, 100012, Beijing, China\\
         \and Graduate University of the Chinese Academy of Sciences, 19A Yuquan Road,
              Shijingshan District, 100049, Beijing, China\\
         \and Sydney Institute for Astronomy (SIfA), School of Physics, The University of Sydney, NSW 2006, Australia\\
         \and Zentrum f\"{u}r Astronomie der Universit\"{a}t
              Heidelberg, Landessternwarte, K\"{o}nigstuhl 12, D-69117 Heidelberg, Germany\\
         \and Division of Astronomy and Space Physics, Department of Physics and Astronomy, Uppsala University, Box 516, 75120 Uppsala, Sweden\\
             }
   \date{}

\abstract{We determined the silicon abundances of 253 metal-poor
  stars in the metallicity range $-4<\mathrm{[Fe/H]} <-1.5$, based
  on non-local thermodynamic equilibrium (NLTE) line formation
  calculations of neutral silicon and high-resolution spectra obtained
  with VLT-UT2/UVES. The $T_{\mathrm{eff}}$ dependence of [Si/Fe] noticed
  in previous investigation is diminished in our abundance analysis
  due to the inclusion of NLTE effects. An increasing slope of [Si/Fe]
  towards decreasing metallicity is present in our results, in agreement
  with Galactic chemical evolution models. The small intrinsic scatter of
  [Si/Fe] in our sample may imply that these stars formed in a region where
  the yields of type II supernovae were mixed into a large volume,
  or that the formation of these stars was strongly clustered, even if the ISM
  was enriched by single SNa II in a small mixing volume. We identified two dwarfs
  with $\mathrm{[Si/Fe]}\sim +1.0$: HE~0131$-$3953, and HE~1430$-$1123. These main-
  sequence turnoff stars are also carbon-enhanced. They might have been
  pre-enriched by sub-luminous supernovae.}


\titlerunning{Chemical evolution of silicon}
\authorrunning{Zhang et al.}

\maketitle

\section{Introduction}

Studying the detailed elemental abundances of metal-deficient stars
in the Galactic halo is a standard approach to probe the origin of
our Galaxy and its early evolution, as many of these stars have
formed from the local counterparts to high-redshift gas clouds during
the early chemo-dynamical evolution of the Galaxy \citep[e.g.][and
reference therein]{bee05}. While abundance ratios as a function of
[Fe/H]\footnote{ [A/B] = $\log(N_{\rm{A}}/N_{\rm{B}}) -
\log(N_{\rm{A}}/N_{\rm{B}})_{\odot}$} provide information about the
chemical enrichment history of the Galaxy, the scatter of these
ratios allow to study mixing processes of the interstellar medium
(ISM) in the early phases of the formation of the Galaxy
\citep[e.g.][] {arg00,kar05a,kar05b}.

In investigations of the enrichment of the ISM, the $\alpha$-elements
(e.g., Mg, Si, Ca, and Ti) are often used as tracer elements, because
their yields depend on the mass and the explosion energy of the SN
and the amount of fallback \citep{kar05b}. Silicon, which is produced
by explosive oxygen burning, belongs to the most abundant metals, and
it can be detected over a wide metallicity range. Besides, some
extreme examples are found, which challenge the enrichment model of
SNe II. For instance, HE~1424$-$0241, an extreme metal-poor star with
$\mathrm{[Fe/H]} = -4.0$, has a very low Si abundance
\citep[i.e., $\mathrm{[Si/Fe]} \sim -1.0$\,dex, ][]{coh07}.
Therefore, Si is an element to probe the enrichment of the ISM.

Previous studies of silicon abundances in metal-poor stars yielded a
range of scatter in [Si/Fe]; typically from $\sim 0.06$\,dex to
0.4\,dex
\citep[e.g.][]{rya96,cay04,coh04,hon04,aok05,pre06,lai08,shi09}.
However, these dispersions can not be simply considered as cosmic
scatter reflecting the ISM mixing process. This is mainly due to
three reasons: (1) the small {\bf sample size of analysis stars} in most of
the above-mentioned studies; (2) when several analyses from the
literature are combined, systematic offsets in the Si abundances due
to different methods of stellar parameter determination and different
structure of model atmospheres may arise, which artificially
increases the scatter in the combined sample; (3) the Si abundance
derived from the \ion{Si}{I} line at 3905\,{\AA}, which is the only
line that can be reliably measured in stars at $\mathrm{[Fe/H]}<-2.5$
may not represent the true value, because this line may be
contaminated by CH lines \citep{cay04} and the abundance determined
from this line shows an abnormal dependence on effective temperature
($T_{\mathrm{eff}}$)\citep{pre06,lai08}. All these may conceal the
``real'' cosmic scatter. Thus, Si abundances determined in a careful
and homogeneous way for a large sample of metal-poor stars are
needed.

Very recently, an NLTE analysis of silicon abundances of metal-poor
stars has been carried out by \citet{shi09}, who discuss the NLTE
effects of the strong \ion{Si}{I} lines at 3905\,{\AA} and
4103\,{\AA}. A strong correlation between the difference of [Si/Fe]
calculated under NLTE and LTE assumptions of these two lines and the
stellar parameters in their sample was noticed. This confirms the
suggestion of \citet{pre06} that Si abundances determined from the
\ion{Si}{I} line at 3905\,{\AA} without NLTE corrections for
metal-deficient star may not be considered as the true values at
$T_{\mathrm{eff}}$ warmer than 5800\,K. From these results, the
anomalous $T_{\mathrm{eff}}$ dependence of [Si/Fe]
\citep{pre06,lai08} can be partially explained. Hence NLTE has to be
taken into account when studying the chemical evolution of Si and the
scatter of [Si/Fe] as a function of [Fe/H].

The aim of this work is thus to obtain detailed silicon abundances of
metal-poor stars, so that the correlation between the abundance
ratios and the stellar parameters and the chemical enrichment of the
ISM are explored. This work is based on spectra of the Hamburg/ESO
R-process Enhanced Star survey (HERES), as described in Section
\ref{Sect:Obs}. The method and the procedures of the abundance
analysis are described in Section \ref{Sect:Analysis}. The results
are presented in \ref{Sect:Results} and discussed in Section
\ref{Sect:Discussion}.

\section{Observations and stellar parameters}\label{Sect:Obs}

The present work is based on the spectra of 253 HERES stars. The
sample selection and observations are described in \citet{chr04}. For
the convenience of the reader, we repeat here that the spectra were
obtained with the Ultraviolet-Visual \'{E}chelle Spectrograph (UVES,
Dekker et al.  2000) mounted on the 8\,m Unit Telescope 2 (Kueyen) of
the Very Large Telescope (VLT). The pipeline-reduced spectra cover
the wavelength range from 3769\,{\AA} to 4980\,{\AA} at a minimum
seeing-limited resolving power of $R=20,000$. The coordinates and
barycentric radial velocities of the stars are listed in Table~1 of
\citet{bar05} (heareafter B05).

We adopt the stellar parameters of B05 in our analysis. In the work
of B05, photometric $T_{\mathrm{eff}}$, metallicity estimated from the
calibration of the \ion{Ca}{II} K-line index along with $B-V$ color
\citep{bee99}, $\log g$ estimated from $\log g - T_{\mathrm{eff}}$
correlation \citep{hon04}, $\xi = 1.8$\,km\,s$^{-1}$, and
$v_{\mathrm{macro}} = 1.5$\,km\,s$^{-1}$ were set as initial guess,
and then were refined in an automated analysis which is based on the
Spectroscopy Made Easy (SME) package by \citet{val96}.  The details
are described in Sections 2 and 3 of B05.


\section{Abundance analysis}\label{Sect:Analysis}

In our analysis, the one-dimensional line-blanked local thermodynamic
equilibrium (LTE) model atmospheres MAFAGS \citep{fuh97}, with
opacity distribution functions (ODF) of \citet{kur92} are employed.
For consistency, solar abundances are the same as B05, i.e., C is
 taken from \citet{all02} and other elements are those of
 \citet{gre98}. During the computation of model atmospheres at
$\mathrm{[Fe/H]} < -0.6$, an $\alpha$-element enhancement of 0.4\,dex
is adopted. A convective efficiency of $\alpha_{\mathrm{mlt}} = 0.5$
is used. For more details on the model atmospheres, we refer the
reader to \citet{gru04}.

\subsection{Line synthesis}

The silicon abundances were determined by spectrum synthesis of the
\ion{Si}{I} lines at 3905.53\,{\AA} and 4102.93\,{\AA}, using the
Spectrum Investigation Utility (SIU) of \citet{ree91}, which computes
line formation under both LTE and NLTE conditions. Continuum scattering
is considered in the computation of the source function.

\cite{shi09} studied the silicon abundance discrepancy between NLTE
and LTE analyses for the two lines adopted in our analysis, and they
suggested that this departure is correlated with the strength of
lines and stellar parameters. The main characteristics are: the NLTE
effects of weak lines is small; the NLTE corrections of these two
lines increase for extremely metal-poor warm stars, and the values
can reach more than 0.15\,dex for the 3905\,{\AA} line and 0.25\,dex
for the 4103\,{\AA} line. Thus, the NLTE effects of these two lines
are considered in the present analysis. The silicon model atom and
the NLTE calculation method are described in detail in
\citet{shi08,shi09}.

Another factor which may affect the determination of the silicon
abundance is contamination with CH lines. \citet{coh04} suggested
that the \ion{Si}{I} line at 3905.53\,{\AA} is probably blended with
the B-X bandhead, which is located approximately at $\lambda =
3900$\,{\AA}.  \citet{pre06} noticed that the blend effect of this CH
band is weak in their sample of red horizontal-branch stars. However,
the [C/Fe] ratio of most of their sample stars is less than 0.0\,dex.
Therefore, in order to get reasonable results for our metal-deficient
sample stars including giants and main-sequence stars, the CH B-X
lines are included in our line synthesis.

Although B05 have already derived the carbon abundance, in order to
keep the consistency of the abundance analysis technique, the
abundance determination for A-X system of CH near 4310\,{\AA} were
independently performed with the analysis code. The oxygen abundance
was adopted to be $\mathrm{[O/Fe]} = 0.6$\,dex.

\begin{figure*}[htbp]
\centering
\includegraphics[width=17cm]{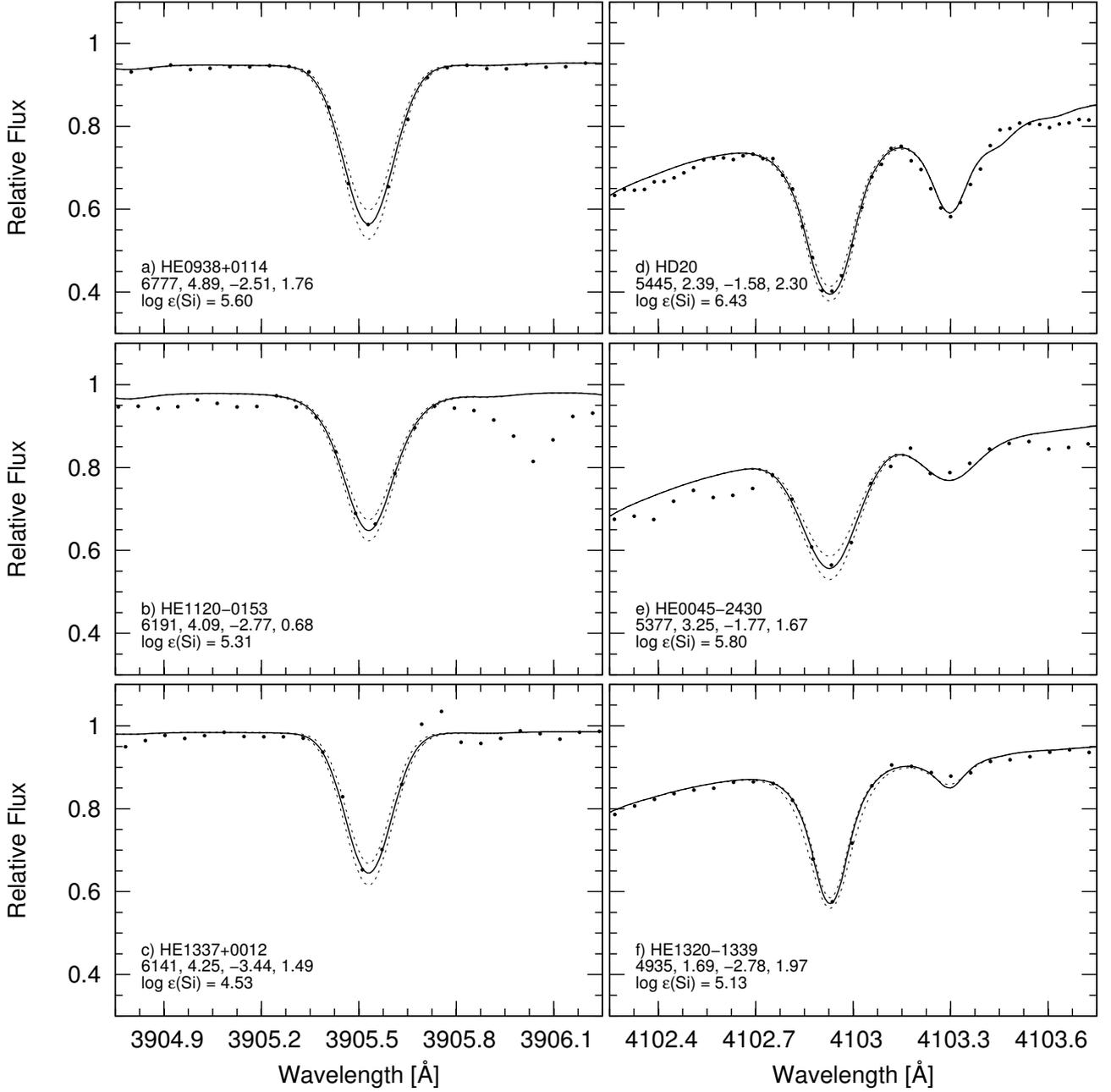}
   \caption{Examples of spectral synthesis for six representative
     stars. The dots are the observational spectra, the solid lines
     are the best-fitting profile, and the dotted lines are the
     synthetic spectra with Si abundances of $\pm 0.15$\,dex relative to
     the best fit, corresponding to less/larger than 5\% in the continuum.
     The listed parameters are $T_{\rm eff}$, $\log g$, [Fe/H], and
      $\xi_{t}$, respectively.  }
      \label{FigVibStab}
\end{figure*}

The atomic line data of \ion{Si}{I} lines are listed in
Table~\ref{Tab:SiLines}. The oscillator strengths ($\log gf$) are
adopted from the experimental results of \citet{gar73}, and van der
Waals interaction constants ($\log \rm C_{6}$, in the unit of $\rm
s^{-1} \rm cm^{6}$, frequency definition) are calculated according
to the interpolation tables of \citet{ans91,ans95}. The molecule line
data of the CH A-X system are taken from B05, and reference therein.
The line positions and $\log gf$ values of the CH lines around
3900\,{\AA} are selected from the database of \citet{kur93}. They are
listed in Table~\ref{Tab:CHLines}. For stars in which neither of the
Si lines can be detected clearly, the feature which is on the position
of theoretical silicon line was fitted, and the {\bf maximum value for
Si that could fit the spectrum} is considered as
the upper limits for the Si abundance. Synthetic spectra for six
representative stars of our sample are shown in Fig. \ref{FigVibStab}.

\begin{table*}[htbp]
\caption{Atomic data of the \ion{Si}{I} lines used in our analysis.}
\label{Tab:SiLines}
\begin{tabular}{ccccc}
\hline\hline
$\lambda$ [{\AA}] & Transition & $E_{\mathrm{low}}$ [eV] & $\log gf$ & $\log C_{6}$ \\
\hline
3905.53 & 3p$^{1}$S$_{0}$ -- 4s$^{1}$P$^{0}_{1}$ & $-$1.909 & $-$1.09 & $-$30.917 \\
4102.93 & 3p$^{1}$S$_{0}$ -- 4s$^{3}$P$^{0}_{1}$ & $-$1.909 & $-$3.14 & $-$30.972 \\
\hline
\end{tabular}
\end{table*}

\begin{table*}[htbp]
\caption{Molecular line data for B-X system of the CH molecule near
3905 {\AA} from \citet{kur93}} \label{Tab:CHLines}
\begin{tabular}{cccc}
\hline\hline
$\lambda$ [{\AA}] & $E_{\mathrm{low}}$ [eV] & $\log gf$ & $\log C_{6}$ \\
\hline
3905.675 & 0.124 & $-$1.178 & $-$32.521 \\
3905.716 & 0.124 & $-$3.862 & $-$32.521 \\
\hline
\end{tabular}
\end{table*}

\subsection{Abundance uncertainties}

The main uncertainties in the abundances are caused by (1)
uncertainties in the analysis of individual lines, including random
errors of atomic data and fitting uncertainties; (2) errors in the
continuum rectification; (3) uncertainties of the stellar parameters.

The errors of $\log gf$ given in \citet{gar73} were adopted as the
perturbation which was added to change the abundance. The variances
of the silicon abundance were taken as the uncertainties affected by
$\log gf$, and they are around 0.02\,dex. It results in an error of
0.02\,dex on average. After getting the best fitting profile of a
certain silicon line, the abundance was changed until the profile
deviates from the best one. This abundance change is adopted as
the fitting uncertainty. Typically, this value is 0.03\,dex, {\bf which is close to the noise.} Finally,
the random error is estimated by summing the estimated error on
the adopted $\log gf$ value and the fitting uncertainty in quadrature.
This result is around 0.04\,dex.

The continuum around the silicon line at $\lambda = 3905.53$\,{\AA}
is affected by the wings of H$\epsilon$ and \ion{Ca}{II} K lines if
the effective temperature exceeds 5500~K in our analysis. It is
difficult to get the accurate continuum location for this wavelength
range in this case, which has a direct effect on abundance
determination for the dwarfs. The situation is similar for the
4102.93\,{\AA} line, which is located in the wing of the H$\delta$
line. In the worst case, the error in continuum rectification was
estimated to be five percent, which results in a change of the Si
abundance of up to 0.11\,dex.

From the determination of atmospheric parameters described in B05,
100\,K, 0.25\,dex, 0.1\,dex, and 0.15\,km\,s$^{-1}$ are the average
uncertainties of $T_{\mathrm{eff}}$, $\log g$, metallicity, and
micro-turbulent velocity, respectively. These uncertainties typically
result in abundance changes of 0.06\,dex, 0.03\.dex, 0.01\,dex, and
0.1\,dex, respectively. The overall uncertainty from errors in the
atmospheric parameters is estimated by summing these four abundance
changes in quadrature.

Finally, the quadratic sum of the uncertainties from these three
sources is adopted as the total abundance error.


\section{Results}\label{Sect:Results}

\subsection{Carbon}

\begin{figure}[htbp]
  \centering
  \includegraphics[scale=0.8]{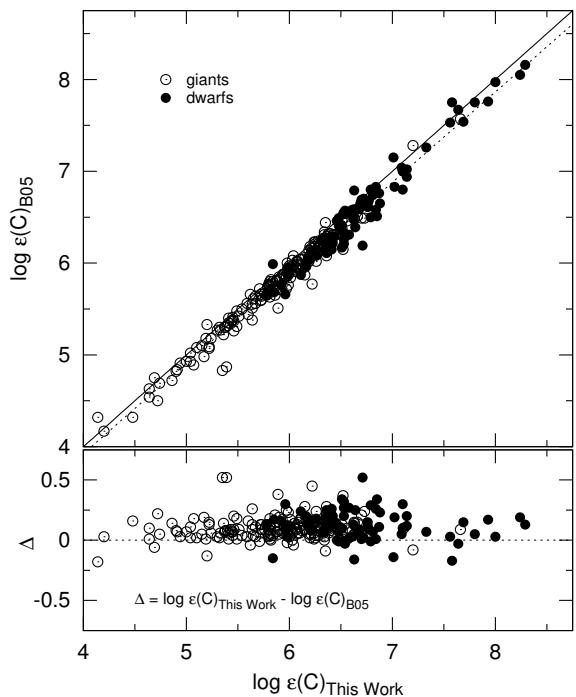}
  \caption{Comparison of the carbon abundance determined in this work
    with those of B05. The open circles refer to giants, while the filled
    circles represent subgiants and dwarfs. The solid line is the
    one-to-one correlation and the dotted line represents a linear
    fit of the data.}
  \label{Fig:logepsC}
\end{figure}

The abundance results are listed in Table~\ref{Tab:Abundances}, and a
comparison with the abundances derived by B05 is shown in Fig.
\ref{Fig:logepsC}. The carbon abundances agree well with each other:

$\log\varepsilon(\rm C)_{\rm B05} = -0.05(\pm 0.07) +
0.99(\pm 0.01) \times \log\varepsilon(\rm C)_{\rm This Work}$

We note that the $\log\varepsilon(\rm C)$ values derived by us are
systematically higher by about 0.10\,dex. This difference can be
explained by the difference of the model atmospheres. The theoretical
continuum computed by the MAFAGS is higher than that calculated by
MARCS (used in B05), which results in a higher carbon abundance.

\begin{figure}[htbp]
  \centering
  \includegraphics[scale=0.85]{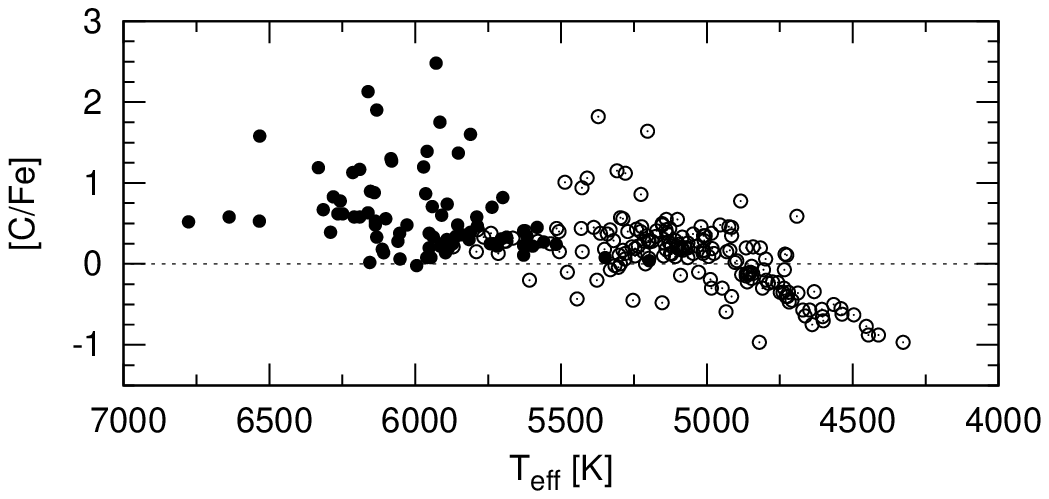}
  \caption{[C/Fe] as a function of $T_{\mathrm{eff}}$. The
    symbols are the same as in Fig.~\ref{Fig:logepsC}.}
  \label{Fig:CFe}
\end{figure}

The carbon abundance ratio as a function of $T_{\mathrm{eff}}$ is
shown in Fig. \ref{Fig:CFe}. The decreasing [C/Fe] towards
decreasing $T_{\rm eff}$ for stars whose $T_{\rm eff}$ are below
5000\,K is {\bf expected}, because the surface abundance of carbon of
evolved giants may be deficient due to the mixing processes
including first dredge-up and extra-mixing \citep{cay04,luc06,
aok07}. For the giants with $T_{\rm eff}$ lower than 5000\,K,
contamination of \ion{Si}{I} 3905\,{\AA} by CH B-X band can be neglected.
Excluding these low temperature giants and
the carbon enhanced stars ([C/Fe] $>$ 1.0~dex \citep[see][]{luc06},
the $<$[C/Fe]$>$ = $0.33\pm0.24$~dex. If the carbon enhanced stars
are accounted in, the average value is changed to 0.42$\pm$0.44~dex,
with larger dispersion. These values imply that the CH B-X band may
{\bf affect} the line profile of \ion{Si}{I} 3905\,{\AA} for most of our
sample stars, thus it is necessary to add CH B-X band in our
line fitting.

\subsection{Silicon}

Our silicon abundance results are also listed in
Table~\ref{Tab:Abundances}. The average value and standard deviation
of the abundance ratios derived by these two lines are as follows:
$<[\rm Si/Fe]_{3905}> = 0.44\pm 0.39$ (247 stars) and $<[\rm
  Si/Fe]_{4103}> = 0.41\pm0.42$ (199 stars). Note that the stars for
which only upper limits are available are not considered in these
calculations.

\begin{figure}[htbp]
  \centering
  \includegraphics[scale=0.8]{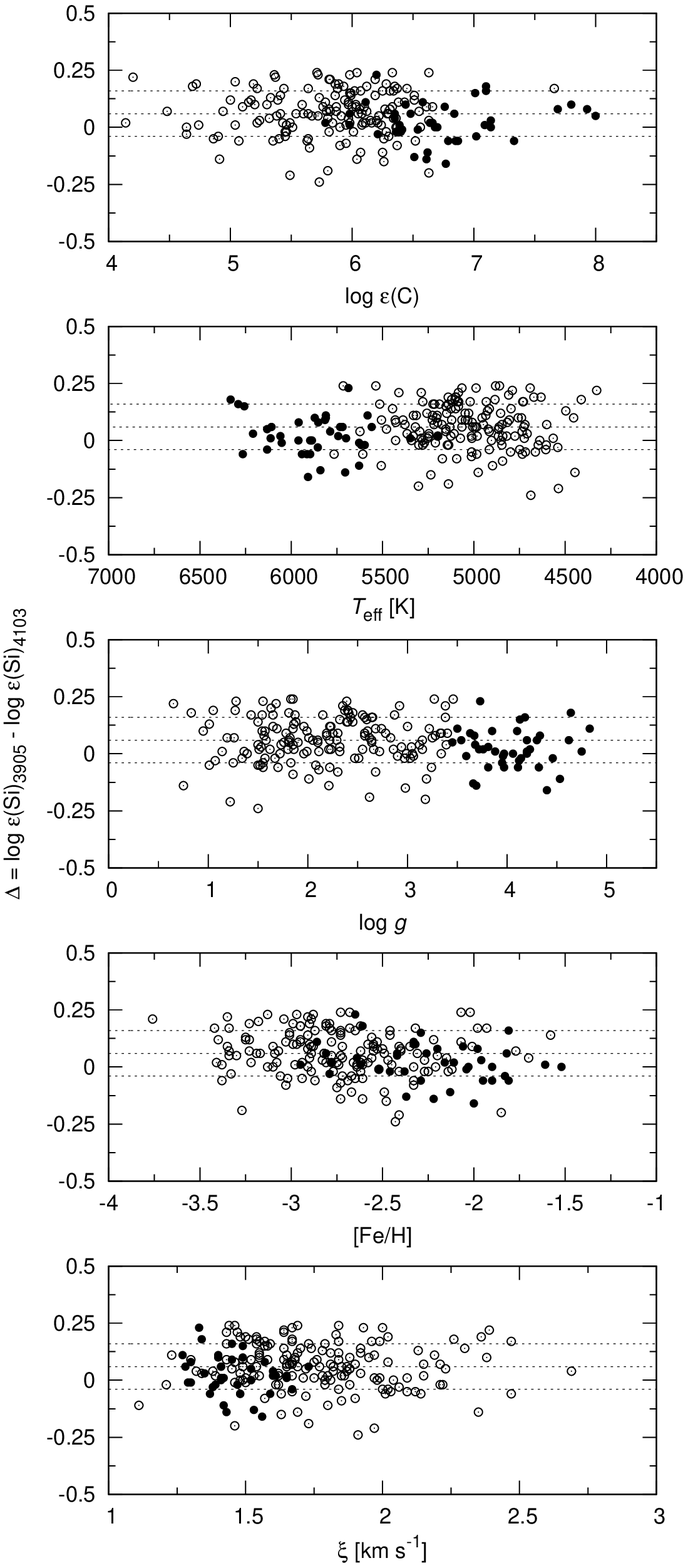}
  \caption{Difference between the abundances of Si determined by the
    \ion{Si}{I} 3905 and 4103\,{\AA} lines as a function of the C
    abundance and stellar parameters. The symbols are the same as in
    Fig.~\ref{Fig:logepsC}. The dashed lines show the average difference
    between these two lines and $1\sigma$ scatter.}
  \label{Fig:DeltaSi}
\end{figure}

The abundance discrepancy between Si$_{3905}$ and Si$_{4103}$ as a
function of the C abundance and the stellar parameters is shown in
Fig.~\ref{Fig:DeltaSi}. In this figure, only the stars that had both
lines measured are used to make a comparison. The dashed lines in
Fig.~\ref{Fig:DeltaSi} present the average difference (0.06\,dex) and
1~$\sigma$ scatter $(\pm0.09$\,dex). In the upper panel one
can notice that there is no trend in $\Delta$ ($ = \log\varepsilon
(\rm Si)_{3905} - \log\varepsilon(\rm Si)_{4103}$) vs.
$\log\varepsilon(\rm C)$. It reflects the fact that the contamination
with the CH B-X band has been eliminated in our final results.

There is a small offset between the results derived from 3905\,{\AA}
and those derived from the 4102.93\,{\AA} line. According to
\citet{shi08}, the 4102.93\,{\AA} line should give a higher abundance
if the $\log gf$ values of \citet{gar73} are adopted
\citep[see][Fig.7]{shi08}. Our results show the contrary. As
discussed above, the blend with CH lines is unlikely to be the
reason. Moreover, most of the sample stars are very metal-poor, thus
blends of other metal components can be neglected. From the panels of
Fig.~\ref{Fig:DeltaSi}, the distribution of the difference shows a
concentration around giants. This phenomenon may be explained by two
reasons:

(1) \citet{lai08} raised the hypothesis that strong lines would lead
to larger abundance values than weak ones, especially in giants, if
the $T-\tau$ relationship of the adopted model atmosphere is
shallower than that of true one. For most of the giants in our
sample, the equivalent width (EW) of the line at 3905\,{\AA}
($\mathrm{EW} > 150$\,m{\AA}) is much larger than that of the
4103\,{\AA} line ($\mathrm{EW} < 120$\,m{\AA}), thus the larger
derived abundance from the 3905\,{\AA} line and a slight increase of
the difference towards decreasing $T_{\rm eff}$ (see the second panel
of Fig.~\ref{Fig:DeltaSi}) are reasonable.

(2) The strong lines are sensitive to the micro-turbulence velocity.
Twenty stars were used as a test: if the $\xi_{t}$ value is increased
0.15\,km/s, the $\log\varepsilon(\rm Si_{3905})$ will decrease by
0.11\,dex, while the $\log\varepsilon(\rm Si_{4103})$ only decreases
by 0.04\,dex. Hence, the determination of $\xi_{t}$ may cause higher
silicon abundances for the 3905\,{\AA} line.

It can also be seen in Fig.~\ref{Fig:DeltaSi} that $\Delta$ decreases
with increasing metallicity. This s probably an artifact caused by the
fact that the 4103\,{\AA} line is difficult to detected at low
metallicity. In these comparison, stars in which only $\rm Si_{3905}$
can be detected are unavailable in such a low metallicity range.

The average of the Si abundance determined from Si$_{3905}$ and
Si$_{4103}$ are taken to represent the final abundance. If only an
upper limit can be derived from one line, we adopt the value derived
from the other line. The average Si abundance ratio and its standard
deviation are $<[\rm Si/Fe]> = 0.46\pm 0.20$ (253 stars).
This value is closed to the prediction in \citet{gos00} ( about
0.5\,dex in the low metallicity regime), while the value is 0.53--0.68\,dex
in the calculation of \citet{kob06}. The higher theoretical value is primarily
due to the adopted IMF in the models, because [$\alpha$/Fe] is higher for larger
stellar masses.

{\bf Considering} the mixing effect in low temperature giants and the accretion from
a companion for the carbon enhanced metal-poor (CEMP) star, an average [Si/C]
of $0.13 \pm 0.21$ in the range
of $0 < \mathrm{[C/Fe]} < 1$ and $T_{\rm eff} >$ 5000\,K was estimated.
In the predictions of \citep{woo95,heg02}, [Si/C] is about 0.15\,dex if the
initial mass of the progenitor star was about 12--40\,M$_{\bigodot}$.

In the upper panel of Fig.~\ref{Fig:SiFe}, we show our results along
with the results of previous LTE silicon abundance analyses. Most of
these studies presented large scatters in [Si/Fe]. For
instance, \citet{rya96} showed that the star-to-star scatter
increases towards decreasing [Fe/H], that is 0.11 for [Fe/H] $ >
-1.5$, 0.14 for $-2.5 <$ [Fe/H] $< -1.5$, and 0.32 for [Fe/H] $\leq
-2.5$. \citet{pre06} gave a star-to-star scatter of 0.22 for 24
giants([Fe/H] $< -2.0$). In our NLTE results, the scatter of dwarfs
is smaller ($\sim 0.13$). Also, for the whole sample, the
star-to-star scatter is close with the estimated uncertainties
($\sim$ 0.16), that is, 0.23\,dex, 0.18\,dex, and 0.16\,dex in the
metallicity ranges of [$-4$,$-3$], [$-3$,$-2$], and [$-2$,$-1$],
respectively. In the lower panel of the same figure, our result
shows stronger correlation between [Si/Fe] and [Fe/H]. The slope of
[Si/Fe] versus [Fe/H] found in our NLTE analysis is $-0.14$ ([Si/Fe]
 = $0.15(\pm0.07) - 0.14(\pm0.03) \times$ [Fe/H]), which is
larger to the values found by most LTE results (e.g., --0.03 in
\citet{mcw95}, --0.07 in \citet{rya96}, 0.03 in \citet{hon04}, --0.06
in \citet{pre06}, and so on). More details are discussed in
Sec.~\ref{Sect:Discussion}.

\begin{table*}[htbp]
\caption{Abundance results of carbon and silicon. The entire table is
  available only electronically. A portion is shown here for guidance
  regarding its form and content. The last column is the average of
  [Si/Fe] from two \ion{Si}{I} lines. If only upper limit can be got
  from one line, taking the value of the other line represents the
  average value.}
\label{Tab:Abundances}
\begin{tabular}{lcccrrrrrrrr}
\hline\hline
     &        &       &        & & \multicolumn{2}{c}{$\log \varepsilon(\rm Si)_{\rm NLTE}$} & \multicolumn{2}{c}{[Si/H]$_{\rm NLTE}$} & ... & ... & ... \\
star & [Fe/H] & $\log \varepsilon(\rm C)$ & [C/H] & [C/Fe] & 3905 & 4103 & 3905 & 4103 & ... & ... & ... \\
\hline
\smallskip
CS22175-007  & --2.81  & 5.80 &  --2.72$\pm$0.14 &   0.09$\pm$0.16  & 5.16 & $<$5.19 & --2.39$\pm$0.13   &  $<$--2.36$\pm$0.15  & ... & ... & ...  \\
CS22186-023  & --2.72  & 6.00 &  --2.52$\pm$0.10 &   0.20$\pm$0.12  & 5.26 &    5.17 & --2.29$\pm$0.09   &     --2.38$\pm$0.11  & ... & ... & ...  \\
CS22186-025  & --2.87  & 5.35 &  --3.17$\pm$0.15 & --0.30$\pm$0.17  & 5.22 &    5.28 & --2.33$\pm$0.14   &     --2.27$\pm$0.16  & ... & ... & ...  \\
CS22886-042  & --2.68  & 5.71 &  --2.81$\pm$0.11 & --0.13$\pm$0.13  & 5.46 &    5.22 & --2.09$\pm$0.10   &     --2.33$\pm$0.12  & ... & ... & ...  \\
CS22892-052  & --2.95  & 6.35 &  --2.17$\pm$0.11 &   0.78$\pm$0.13  & 5.31 &    5.13 & --2.24$\pm$0.10   &     --2.42$\pm$0.12  & ... & ... & ...  \\
. & . & . & . & . & . & . & . & . & . & . & . \\
. & . & . & . & . & . & . & . & . & . & . & . \\
. & . & . & . & . & . & . & . & . & . & . & . \\
HE2338-1618  & --2.65  & 6.31 &  --2.21$\pm$0.10 &   0.44$\pm$0.12  & 5.41 &    5.25 & --2.14$\pm$0.09   &     --2.30$\pm$0.11  & ... & ... & ...  \\
HE2345-1919  & --2.46  & 6.40 &  --2.12$\pm$0.10 &   0.34$\pm$0.12  & 5.58 &   5.60 & --1.97$\pm$0.09   &     --1.95$\pm$0.11  & ... & ... & ...  \\
HE2347-1254  & --1.83  & 7.02 &  --1.50$\pm$0.14 &   0.33$\pm$0.16  & 6.07 &   6.11 & --1.48$\pm$0.13   &     --1.44$\pm$0.15  & ... & ... & ...  \\
HE2347-1334  & --2.55  & 5.20 &  --3.32$\pm$0.13 & --0.77$\pm$0.15  & 5.36 &   5.26 & --2.19$\pm$0.12   &     --2.29$\pm$0.14  & ... & ... & ...  \\
HE2347-1448  & --2.31  & 6.84 &  --1.68$\pm$0.11 &   0.63$\pm$0.13  & 5.79 & $<$5.74 & --1.76$\pm$0.10   &  $<$--1.81$\pm$0.12  & ... & ... & ...  \\
\hline
\end{tabular}
\end{table*}


\section{Discussion and conclusions}\label{Sect:Discussion}

\subsection{Abundance correlations with stellar parameters}\label{Sect:avp}

\begin{figure}[htbp]
  \centering
  \includegraphics[scale=0.85]{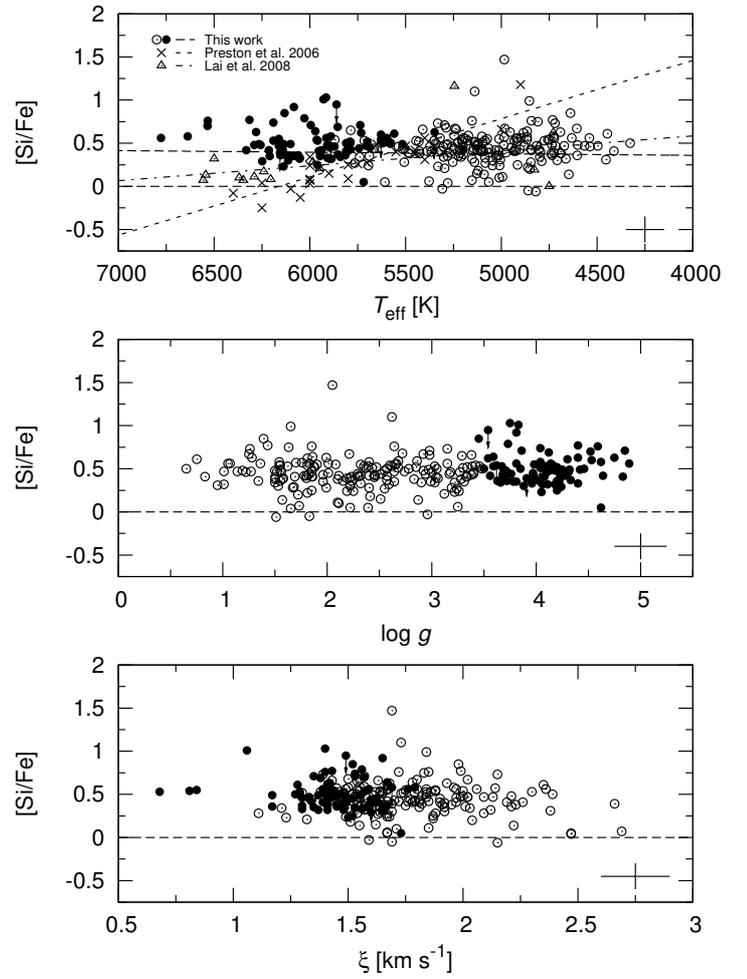}
  \caption{Si abundance ratio as a function of stellar parameters.
    The arrows refer to upper limits; otherwise, the symbols are the
    same as in Fig.~\ref{Fig:logepsC}. The average error bar is shown
    in the lower right corner of each panel. The crosses are the
    results of \citet{pre06}, the while the open triangles are the
    ones of \citet{lai08}. Besides, in the upper panel, dashed line,
    short dashed line, and dot dashed line represent the least square
    fits of the results of our observed data, \citet{pre06}, and
    \citet{lai08}, respectively.}
  \label{Fig:SiFeStellarParameters}
\end{figure}

\begin{figure}[htbp]
  \centering
  \includegraphics[scale=0.85]{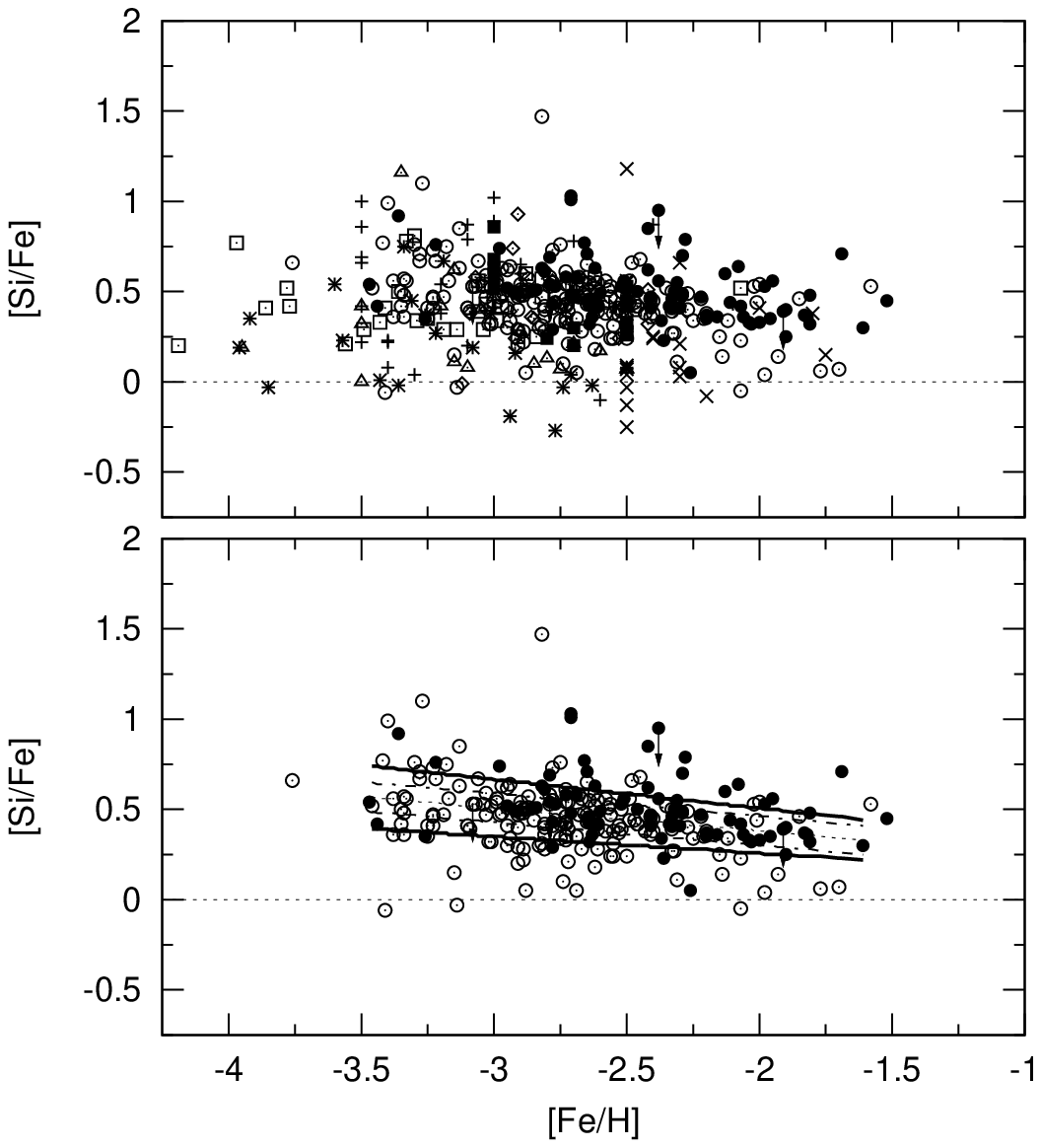}
  \caption{Same as Fig.~\ref{Fig:SiFeStellarParameters}, but for
    [Si/Fe] vs. [Fe/H]. In the upper panel, we compare
    our results with those of previous LTE analyses:
    \citet[][pluses]{mcw95}; \citet[][asteriks]{rya96}; \citet[]
    [open square]{cay04}; \citet[][open diamonds]{hon04};
    \citet[][filled square]{aok05}. In the lower panel, thick solid lines are
    1-$\sigma$ scatter, short dashed lines crossing data points represents
    the fitting slope, and dashed-dotted lines are the fitting uncertainties.}
      \label{Fig:SiFe}
\end{figure}

In Figs.~\ref{Fig:SiFeStellarParameters} and \ref{Fig:SiFe}, we show
[Si/Fe] as a function of the stellar parameters. The abundance
correlation with stellar parameters is discussed below.

Previous LTE silicon abundance analyses of metal-poor stars reported
a correlation of [Si/Fe] with $T_{\mathrm{eff}}$
\citep[e.g.][]{pre06,lai08}, i.e., [Si/Fe] decreases with increasing
temperature. In our results with NLTE correction, the phenomenon is
not obvious. The slopes of these three data sources are listed below:

(1) This work:

[Si/Fe] = $0.29(\pm0.13) + 0.33(\pm0.23) \times T'_{\rm eff}$

(2) \citet{pre06}:

[Si/Fe] = $4.16(\pm0.39) - 6.74(\pm0.68) \times T'_{\rm eff}$

(3) \citet{lai08}:

[Si/Fe] = $1.28(\pm0.48) - 1.74(\pm0.83) \times T'_{\rm eff}$

Note that $T_{\rm eff} = T'_{\rm eff} \times 10^{4}$.

These relationships can be also seen in the upper panel of
Fig.~\ref{Fig:SiFeStellarParameters}, in which our results are
plotted along with previous LTE abundance analyses. The steep slope
in [Si/Fe] versus $T_{\mathrm{eff}}$ in previous studies is mainly
caused by the low [Si/Fe] stars hotter than $\sim 5500$\,K. The NLTE
correction decreases with decreasing temperature. At higher
$T_{\mathrm{eff}}$, the results with NLTE correction will become
larger, which causes a higher silicon abundance than those of LTE and
makes this slope much smaller. Therefore, our results support the
conclusion of \citet{shi09} that NLTE effects can explain the
temperature dependency of [Si/Fe].
Therefore, the increasing trend of [Si/Fe] with the declined
$T_{\mathrm{eff}}$ is diminished, if NLTE is
considered in the abundance analysis of silicon.

\citet{pre06} concluded that there was no correlation between [Si/Fe]
and $\log g$, and our NLTE results also confirm this conclusion.

In the lower panel of Fig.~\ref{Fig:SiFe}, an increase of [Si/Fe]
with decreasing [Fe/H] can be seen. Although \ion{Fe}{I} is affected
by significant NLTE effects for giants and very metal-poor stars
\citep[e.g. $T\mathrm{eff} = 5000$\,K, $\log g = 2.00$, and
$\mathrm{[Fe/H]} = -3.00$,][]{mas10}, the NLTE correction of
\ion{Fe}{I} leads only to small changes in our final [Si/Fe] results
and the slope of [Si/Fe] vs. [Fe/H]. In the worst case, we find a
NLTE correction for [Fe/H] of $+0.25$\,dex, corresponding to a change
in [Si/Fe] of $+0.03$\,dex. Applying the corrections to our 22 very
metal-poor giants ($\mathrm{[Fe/H]} < -3.0$\,dex) would lead a change
of $+0.02$ in the slope of [Si/Fe] vs. [Fe/H]. In addition, the
corrections for stellar granulation for Si and Fe are small (i.e., $<
0.1$\,dex), and significant only for high-excitation potential lines
in metal-deficient stars \citep{asp05b}. Therefore, we conclude that
the observed slope in Fig.~\ref{Fig:SiFe} may not be the result of
NLTE/3D effects.

\begin{figure}[htbp]
  \centering
  \includegraphics[scale=0.85]{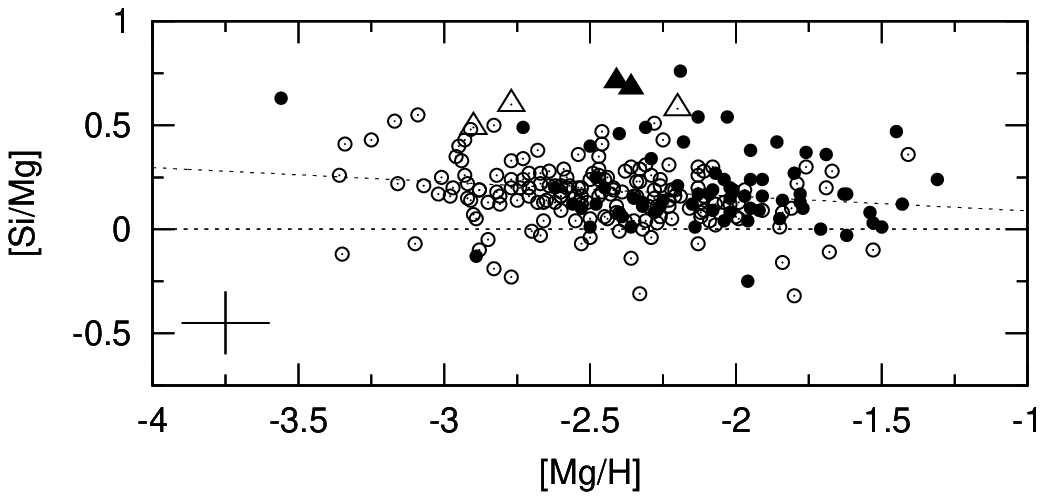}
  \caption{[Si/Mg] as a function of [Mg/H]. The symbols are the same
    as in Fig.~\ref{Fig:SiFeStellarParameters}. The big triangles are
    marked as the Si-enhancement stars. Open ones are giants, while
    filled ones are dwarfs.}
  \label{Fig:SiMg}
\end{figure}

Magnesium is also used as the tracer to discuss the metallicity
dependence. In Fig.~\ref{Fig:SiMg}, [Si/Mg] against [Mg/H] is
plotted, where the magnesium abundances are taken from B05. A slope
of [Si/Mg] vs. [Mg/H] can be noticed: [Si/Fe] = $0.02(\pm0.06) - 0.07
(\pm0.03) \times$ [Mg/H]. The NLTE effect of Mg may not be
the reason which causes this tendency. This is because in the very recent
NLTE study of Mg of \citet{and10}, the NLTE results of Mg have the
same evolution behavior as the LTE ones, and the NLTE correction of
Mg just enhances the abundance.

More discussion about the trends will be presented in \ref{Sect:mixing}.

\subsection{The outliers in our sample}\label{Sect:out}

We did not find any stars with a deficiency of Si \citep[such as
HE~1424$-$0241,][]{coh07}. This star is at [Fe/H]$\sim -4$ with an
unusually low Si abundance such that [Si/Fe]$= -1.01$
and[Si/Mg]$=-1.45$. \citet{coh08} speculated that this phenomenon may
be the result of a chemically inhomogeneous ISM and that the star
probably was enriched by a single SN. If so, our results imply that
our sample stars may not be formed in the gas which was contributed
by ejecta from only one SN. This will be discussed further in
Sec.~\ref{Sect:model}.

On the other hand, we noticed five candidates with large
overabundance of silicon, [Si/Fe] are 1.47\,dex, 0.99\,dex,
1.10\,dex, 1.01\,dex, and 1.03\,dex for HE~0308$-$1154,
HE~1246$-$1344, HE~2314$-$1554, HE~0131$-$3953, and HE~1430$-$1123,
respectively. The first three are giants and the other two are dwarfs. Only
HE~0308$-$1154 whose [Si/Fe] is outside of the 3$\sigma$ limit can be
clearly considered as Si-enhancement (in our observed sample, [Si/Fe]
is in Gaussian distribution, that is $\# = 253., \mu = 0.46, \sigma = 0.20$).
To probe the nature of these stars, we investigate the abundance patterns
of these stars, as derived by B05, and discuss them below.

\textbf{Giants:}

Two additional metal-deficient giants with large Si-enhancement are
known:

(1) CS29498$-$043 [Fe/H]=$-$3.75\,dex, [C/Fe]=1.90\,dex,
[Mg/Fe]=1.81\,dex, [Si/Fe]=1.07\,dex \citep{aok02}

(2) CS22949$-$037 [Fe/H]=$-$3.79\,dex, [C/Fe]=1.05\,dex,
[Mg/Fe]-1.22\,dex, [Si/Fe]=1.04\,dex \citep{nor01}.

Both of them are CEMP stars with a large excess of $\alpha$-elements.

However, in our study, the giants HE~0308$-$1154, HE~1246$-$1344, and
HE~2314$-$1554 have otherwise ``normal'' abundance ratios. We checked
the EW of two \ion{Si}{I} lines of these three stars, and found that
both of the EWs of these lines are larger than 100 m{\AA}, and the
differences of derived abundance between \ion{Si}{I} 3905 and 4103
are small. The incorrect "T-$\tau$" relationship in model atmosphere
\citep{lai08} can results in an offset of 0.2\,dex. This phenomena
can be partially interpreted by the following hypothesis.

\textbf{Dwarfs:}

Previously, large excesses of Si were rarely found in dwarfs. The
[Si/Fe] value of metal-deficient dwarfs determined by using
\ion{Si}{I} transitions in the red spectral region which are not
affected by NLTE effects, are seldom higher than 0.6\,dex
\citep[e.g.][]{ste02,shi09,zha09}, but these lines are difficult to
detected at $\mathrm{[Fe/H]}<-2.0$\,dex. Even assuming a NLTE
correction of $+0.2$\,dex for the [Si/Fe] values determined by
\citet{pre06,lai08}, where the Si abundance is derived from the
3905.93\,{\AA} line, none of the stars in their sample would be
Si-enhanced by more than 0.75\,dex.

The two Si-enhanced dwarfs, HE~0131$-$3953 and HE~1430$-$1123, are
Ba-enhanced CEMP stars. Furthermore, HE~0131$-$3953 was identified as
an s-II star \footnote{this kind of star is also called $r + s$ star
\citep{jon06}} by B05, and HE~1430$-$1123 has rather low [Sr/Ba]
value of $-1.58$\,dex, which is thought to be associated with the
s-II stars. This star can not be identified as a s-II star because of
lacking abundance information for Eu (see more details in B05).
Although mass transfer from a formerly more massive companion during
its AGB phase might have caused the enhancements of C and Ba seen in
these stars, this scenario does not provide an explanation for the
Si-enhancements. \citet{tsu03} suggested that it might be due to
pre-enrichment by subluminous SNe experiencing mixing and fallback.
The fallback which occurred inside the Si layer in subluminous SNe
can result in smaller abundances of
elements heavier than Si and the enhancement of Si in these CEMP
stars relative to iron and the abundance ratio in the Sun.

\subsection{Star-to-star scatters and mixing of the interstellar medium}\label{Sect:mixing}

The dispersion in the abundance ratios of metal-poor stars provides a
measure of the chemical inhomogeneities in the star-forming gas, and hence
of the mixing processes in the ISM. \citet{aud95} argued
that increasing inhomogeneity is to be expected with decreasing metallicity,
as a result of the small number statistics of enriching events (i.e., SN II).
This was also observed for a number of element ratios \citep{rya96,mcw97}.

In the wake of these findings, \citet{arg00} derived the expected scatter for
several abundance ratios, including [Si/Fe], as a function of
metallicity. They predict a star-to-star scatter of $\sim 0.4$\,dex in [Si/Fe]
in the range of $-4 <\mathrm{[Fe/H]} < -3$, at which the model ISM was
essentially unmixed. The scatter reduces to $\sim 0.25$\,dex in the range
$-3 < \mathrm{[Fe/H]} < -2$ due to a gradually increased mixing. At
$\mathrm{[Fe/H]}>-2.0$, the scatter is around $0.2$\,dex, reaching typical
levels of the observational uncertainties depending on the data quality.

In contrast, more recent studies have reported on a number of elements
for which the scatter in the abundance ratios, like [Mg/Fe], are
consistent  with the observational uncertainties, all the way down to
[Fe/H] $\sim -3.5$ \citep[e.g., B05;][]{coh04,arn05,lai08,bon09}. In the present
study, the 1-$\sigma$ scatter in [Si/Fe] is 0.23\,dex, 0.18\,dex, and 0.16\,dex in
the metallicity range [$-4$,$-3$], [$-3$,$-2$], and [$-2$,$-1$], respectively.
Because the halo ISM should be well mixed at metallicities higher than
$-2.0$\,dex, as suggested by minimal mixing models like the one by \citet{arg00},
the scatter of 0.16\,dex can be considered as the observational error.
If so, the cosmic scatter is less than 0.15\,dex in the full range $-4 < \mathrm{[Fe/H]}<-2$,
which is considerably smaller than what was predicted by \citet{arg00}. It therefore seems
that also Si belongs to the class of elements that show very little cosmic scatter.
However, extreme outliers do exist also in [Si/Fe] \citep[see][]{coh07}.
It is not entirely known which role such outliers play. Have they been formed
out of gas enriched by SNe in a specific mass range or are they ``freak objects''
formed under very particular circumstances? In the latter case, the measured
surface abundances may not uniquely reflect common SN nucleosynthesis.
We shall further discuss these issues in the next sections.

\subsubsection{Stochastic modelling of the chemical evolution of Si}\label{Sect:model}

\begin{figure}
\centering
\includegraphics[scale=0.32]{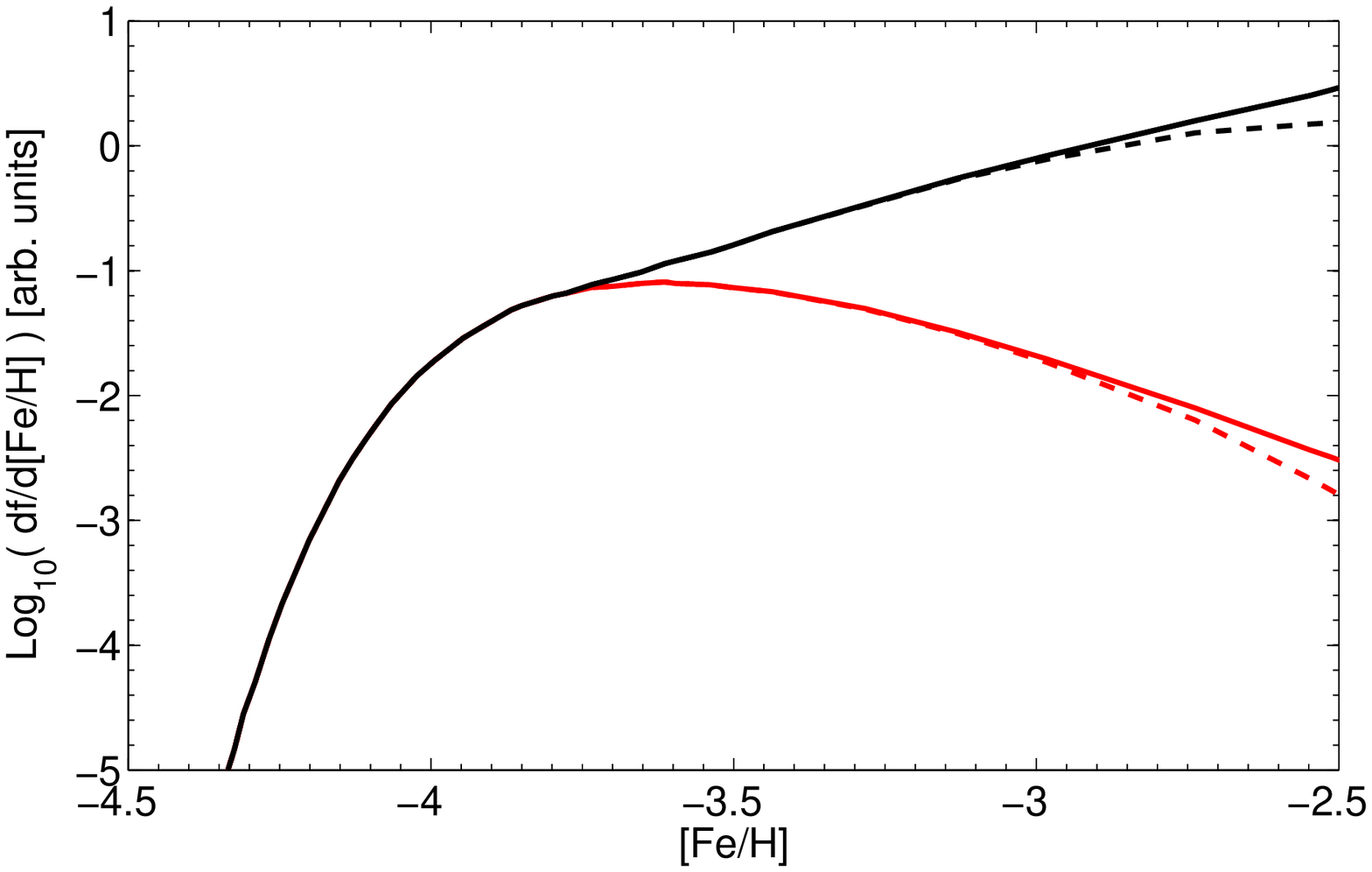}
\caption{The logarithm of the predicted metallicity distribution
function (MDF). The quantity f is the fraction of stars that fall
within each [Fe/H] bin (1 dex). The black, solid line shows the
metal-poor tail of the predicted MDF of the Galactic halo while the
black, dashed line shows the predicted MDF of our observational
sample. The red, solid and dashed lines denote the distribution of
stars enriched by a single SN for the Galactic halo and the current
sample, respectively. Below [Fe/H]$\sim -3.8$, the number of stars
quickly goes to zero.} \label{mdf}
\end{figure}

 In order to investigate the enrichment and amount of mixing in the early
ISM, our large, homogeneous sample is compared with a stochastic model of the
chemical evolution of Si. The statistics discussed here are based on a
model originally developed by \citet{kar05b,kar06} and \citet{kar08}. In this
model, stars are assumed to form randomly within the system. They enrich their
surroundings locally, by ejecting heavy elements such as Si and Fe. The Fe
yields used to calculate the metallicity distribution function (MDF) depicted in
Fig. \ref{mdf}, are taken from \citet{ume02}, which are nearly identical to the
Fe-yields presented in \citet{nom06}. The turbulent mixing of the ISM is
modeled as a diffusion process such that each individual SN remnant
continues to grow in time as

\begin{equation}
V_{\mathrm{mix}}(t) =  \frac{4\pi}{3}(6D_{\mathrm{turb}}t +
\sigma_{E})^{3/2}, \label{vmix}
\end{equation}

\noindent where $V_{\mathrm{mix}}$ is the mixing volume and
$D_{\mathrm{turb}}=1.2\times 10^{-4}$ kpc Myr$^{-1}$ is the turbulent
diffusion coefficient. Here, $\sigma_{E}$, which is a measure of the
initial size of the SN remnant as it merges with the ambient medium,
is set to zero.The model used to calculate the MDF is nearly identical
to model A in \citet{kar05b}.

The large number of stars in the present sample enables us to discuss
outlier statistics. For example, what is the probability of finding an extreme
Si abundance star, similar to HE~$1424-0241$ \citep{coh07}, in our sample?
We shall make the simplifying assumption that stars with such extreme
[Si/Fe] ratios can only occur if they were enriched by a single SN
\citep{coh08} within a certain range of masses. Theoretically, the
low Si-star may be enriched by two, or more SNe,
all within that same mass range but this probability quickly goes to
zero if the fraction of SNe within this range is $\lesssim30\%$, or
so.  About $16\%$ of all Galactic halo stars are found to have a metallicity
below [Fe/H]$=-2.5$ \citep{car96}. Assuming that stars enriched by one SN
predominantly are found in this metallicity regime (see Fig. \ref{mdf}), the
probability of finding a star enriched by a single SN in the Galactic
halo is thus estimated to $p_{1,\mathrm{halo}} = 9\times 10^{-3}$, given
the simulated metallicity distribution function (MDF) in Fig. \ref{mdf}.

As our sample is biased against stars above [Fe/H]$\sim -2.5$, this
must be accounted for if we seek to directly compare the observations
with the model. A selection function of $B-V=0.7$ was adopted
\citep[see][their Table. 12]{sch09}.  While stars enriched by one SN
are hardly affected at all by this bias (\i.e., they are mostly found below
[Fe/H]$=-2.5$), the number of stars enriched by more than one SN is
significantly smaller, by a factor of $\sim 7$. Consequently, the
fraction of stars enriched by single SNe in the present observational
sample is higher, as compared to the corresponding fraction of the
Galactic halo (see Fig. \ref{mdf}). The biased fraction is estimated
to $p_{1,\mathrm{bias}} = 6.1\times 10^{-2}$.

The probability of finding exactly $k$ stars with similarly extreme
abundances like HE~$1424-0241$, in a sample of
$n$ stars is given by the Binomial statistics $B(n,k) = C(n,k)
p^kq^{n-k}$, where $p$ is the probability of success, $q=1-p$ and
$C(n,k) = n!/k!(n-k)!$. Given that only a fraction, $f_{\mathrm{xtrm}}$,
of the stars enriched by a single SN may show an extreme abundance,
the probability of finding such a star is therefore
$p_{\mathrm{xtrm}} = f_{\mathrm{xtrm}}\,p_{1,\mathrm{bias}}$.  The
fraction $f_{\mathrm{xtrm}}$ depends critically on the stellar yields and
the IMF. Both parameters are uncertain, in particular in this extremely
metal-poor regime.

\subsubsection{Abundance ranges, dispersions and outlier statistics}

Including the low Si-star HE~$1424-0241$, the observed range in
[Si/Fe] between this star and the mean of the sample is $\sim1.5$ dex.
The lowest $33\%$ of this range, will still keep us below [Si/Fe]$ =
-0.5$ (i.e., outside $\sim 5\sigma$ of the current sample), which is $\ge0.5$
dex below the next lowest observed [Si/Fe]
ratio at $ \sim 0$. From current observations, we are unable to
estimate how big $f_{\mathrm{xtrm}}$ is in this lower range. However,
even though the theoretical yields do not predict such low values in
[Si/Fe], we can estimate $f_{\mathrm{xtrm}}$ by calculating the
fraction of stars that falls within the lowest $33\%$ of the
corresponding theoretical range. This range, as predicted by the
yield calculations of \citet{heg08}, is reached by $7.5\%$ of the
massive stars within $10 - 40~M_{\odot}$, for a Salpeter IMF. The
corresponding fraction using the yields by \citet{nom06} is $41.5\%$,
in the mass range $13 - 40~M_{\odot}$. We will adopt a fiducial value
of $f_{\mathrm{xtrm}} = 0.15$, and allow for a range of $0.05\le
f_{\mathrm{xtrm}}\le 0.45$.

The probability of finding one or more stars ($k\ge 1$) with a low
[Si/Fe] in a sample of $n=253$ stars can be expressed as
$B(n=253,k\ge 1) = 1 - (1-p_{\mathrm{xtrm}})^{n}  = 90.2\%$,
in the case of $f_{\mathrm{xtrm}}=0.15$ and
$p_{1,\mathrm{bias}} = 6.1\times 10^{-2}$.  Within the range
$f_{\mathrm{xtrm}} = 0.05 - 0.45$, the chance is $B =  53.8 - 99.9\%$,
with increasing B for increasing $f_{\mathrm{xtrm}}$. This is high, irrespectively of
the value of $f_{\mathrm{xtrm}}$. For $f_{\mathrm{xtrm}}=0.075$, the
chance is $B=68.7 \simeq 70\%$.  Hence, the probability is high that at
least one star with an extremely low [Si/Fe] would have been detected in
the current sample. However, as noted in Sect. \ref{Sect:out}, there are no
such stars in our sample.  In this respect, our observations appear
inconsistent with an inhomogeneous ISM in which the metal-poor
stars in the Galactic halo were enriched only by a small number of
SNe, as indicated by the presence of HE~$1424-0241$ at [Si/Fe]$=-1.01$.
The fact that the star found by \citet{coh07} have such a low [Si/Fe]
and appears so detached from the rest of the halo stars, which all
have [Si/Fe]$\gtrsim 0$, may suggest that its Si abundance is not
(only) a result of enrichment by regular core collapse SNe (cf.
\citep{coh07}). If so, we should exclude it from
the comparison between the observed and simulated star-to-star
scatter. This view is also supported by the findings above that more
such stars would likely have been detected in our sample if this star
was a ``normal'' outlier,  enriched by a regular core collapse SN.


In what follows, we shall exclude HE~$1424-0241$ in the discussion
and only consider the sample stars presented here (Table \ref{Tab:Abundances}).
Consequently, the observed range in [Si/Fe] is significantly reduced,
with a star-to-star scatter of $\sigma=0.22$ below [Fe/H]$=-3$. As illustrated in
the upper panel of Fig. \ref{Fig:SiFe_sim}, the observed 1-$\sigma$ scatter is
comparable to the theoretical dispersions expected from the yield
ratio of Si-to-Fe over the mass range of core collapse SNe (the
distributions in Fig. \ref{Fig:SiFe_sim} are convolved with a gaussian
($\sigma=0.14$), to account for the random errors in the observations).
The yield calculations by \citet{nom06} infer a dispersion of $\sigma=0.33$
while the calculations by \citet{heg08} infer a dispersion of
$\sigma=0.23$, or $\sigma=0.27$, if the full mass range
$10\le m/\mathcal{M_{\odot}}\le 100$ is considered. Moreover, the
observed range, [Si/Fe]$_{\mathrm{max}}$ -- [Si/Fe]$_{\mathrm{min}}=1.53,$ is
larger than the expected, theoretical range predicted by Heger \& Woosley
(2008\nocite{heg08}, the observed range is larger in $>99.9\%$ of the
cases for $n=253$ stars, assuming SN progenitor masses in the range
$10\le m/\mathcal{M_{\odot}}\le 40$), while it is comparable to the one
predicted by \citet{nom06}, larger in $38\%$ of the cases).

Note that these are the maximum theoretical dispersions and ranges. In
reality, we expect the stars enriched by a single SNe to be distributed
over a range in [Fe/H]. In particular, a fraction of the stars below [Fe/H]$= -3$
are expected to be enriched by more than one SN. These stars are closer to
the mean [Si/Fe] and the observed 1-$\sigma$ dispersion below [Fe/H]$= -3$
(see Fig. \ref{Fig:SiFe_sim}, lower panel) is therefore expected to be lower
than the dispersion of the yield ratio depicted in Fig. \ref{Fig:SiFe_sim},
upper panel. The lower panel of Fig. \ref{Fig:SiFe_sim} shows a simulation in which
the turbulent mixing is turned off (i.e., minimal mixing). The size (mixing volume) of
the SN remnants is set to $\sigma_{\mathrm{E}}=8.5\times 10^{-3}$, which corresponds
to a mixing mass of $1\times 10^5~\mathcal{M_{\odot}}$, for a particle density of
1cm$^{-3}$. The SN II yields are taken from \citet{nom06}. Apart from the overall
trend, which is shallower in the simulation, the 1-$\sigma$ scatter in the metallicity
three bins $[-4,-3]$, $[-3,-2]$, and $[-2,-1]$, are found to be 0.23, 0.16, and 0.14,
respectively, excluding the stars predominantly enriched by electron capture SNe
(see below). This is significantly smaller than the scatter predicted by \citet{arg00}
and in close agreement with observations. Since we have turned off the turbulent
mixing in our simulations, the discrepancy between the two model results should
predominantly be due to differences in the adopted SN yields.

In conclusion, we cannot reject the possibility that the stars in our
sample were formed in a chemically inhomogeneous ISM, solely based on the
measurements of Si. Admittedly, our sample lack extremely Si-deficient
stars, but this may rather suggest that HE~$1424-0241$ is very atypical, and
should not be included in the analysis. If this star was born with such a low
Si abundance, reflecting the nucleosynthesis of a rare SN, the early ISM must, indeed,
have been highly inhomogeneous. Note that gas that low in Si will rapidly reach
``chemical normality'' as soon as SNe II enrich it. Low Si-stars would therefore
be relatively uncommon. Moreover, the observed scatter increases faster with
decreasing [Fe/H] than does the mean observational uncertainty of the stars
(see Fig. \ref{Fig:SiFe}, lower panel). This
suggests that the scatter at the lowest metallicities has a small but non-negligible
contribution from real abundance inhomogeneities in the early star-forming gas.

\subsubsection{Contribution from electron capture SNe}

\begin{figure}
\begin{center}
  \includegraphics[scale=0.5]{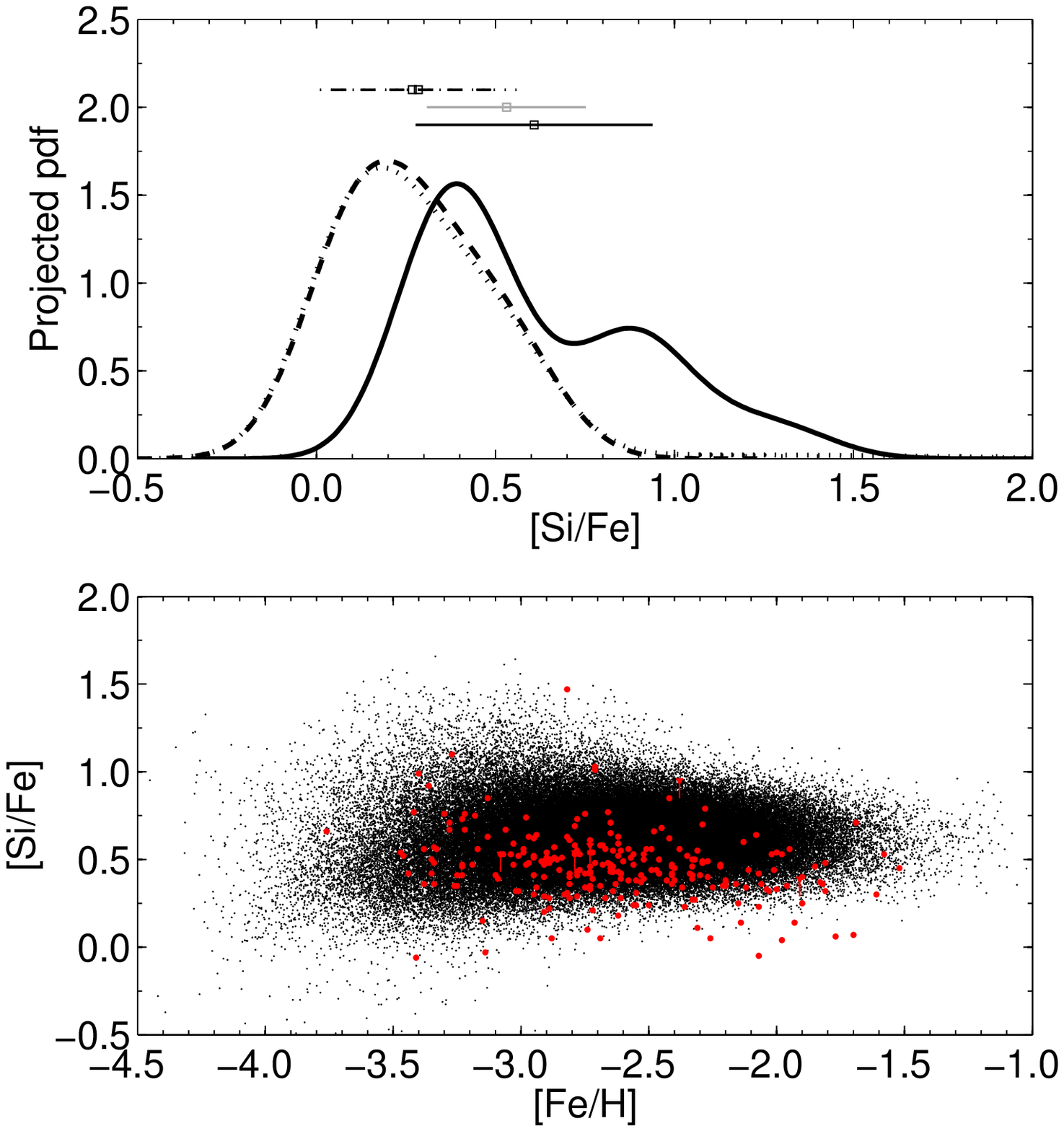}\\
  \caption{The expected star-to-star scatter in [Si/Fe] for core collapse SNe. The top panel shows the expected maximum range in [Si/Fe] for stars enriched by single Type II SNe. The solid curve denotes the probability density function (PDF) assuming SN yields by Nomoto et al. (2006) while the dashed curve (SN mass range $10\le m/\mathcal{M_{\odot}}\le 40$) and the dotted curve ($10\le m/\mathcal{M_{\odot}}\le 100$) denote the PDFs assuming yields by Heger \& Woosley (2008). Each PDF is convolved with a gaussian ($\sigma=0.14$) to account for the observational uncertainty in [Si/Fe]. The corresponding 1-$\sigma$ dispersions are shown as solid ($\sigma=0.33$), dashed ($\sigma=0.23$), and dotted ($\sigma=0.27$) thin lines, centered at the respective mean of each distribution. The gray thin line at [Si/Fe] = 0.53 denotes the observational star-to-star scatter ($\sigma=0.22$) below [Fe/H]$=-3$. The bottom panel shows the full (convolved) distribution of model stars (small black dots) in the [Fe/H] -- [Si/Fe] plane. The observations are shown as red dots (upper limits are shown as triangles), for comparison.  The star-to-star scatter below [Fe/H]$=-3$ in the simulation is $\sigma\simeq 0.23$. Note the small number of EMP stars below [Si/Fe]$\sim 0$. These stars have partly been enriched by electron capture SNe in the mass range $8\le m/\mathcal{M_{\odot}} \le 10$, which produce very small amounts of Si. The SN II (Fe-core collapse) yields are taken from \citet{nom06} while the electron capture SN yields are taken from \citet{wan09}. }
  \label{Fig:SiFe_sim}
\end{center}
\end{figure}

To find out the frequency of low-Si stars enriched by a rare type of SNe, we
included the contribution of electron capture SNe, which have masses in the
range $8-10~\mathcal{M_{\odot}}$. The electron capture SNe are believed to
constitute a fraction of $\sim 4-30\%$ of all SNe \citep{pol08,wan09,wan10}. During
the final stage of their evolution, these objects develop a degenerate
O-Ne-Mg core and their structure and nucleosynthesis are distinctly different
from the more massive Fe-core collapse SNe, including a very low Si yield
\citep{wan09}. Assuming that all stars in the mass range $8-10~\mathcal{M_{\odot}}$
become electron capture SNe (i.e., $30\%$ of all SNe, given a Salpeter IMF), the
fraction of stars in the simulation with a [Si/Fe]$<-0.5$ is
$p_{\mathrm{xtrm}}\simeq1.55\times 10^{-3}$. This gives a probability of
$32.4\%$ of finding a low [Si/Fe] star in our sample. It is a relatively low
probability but not extremely low, and the possibility to find such a star in
the combined sample of Galactic halo stars studied with detailed spectroscopy
is non-negligible. The lower panel of Fig. \ref{Fig:SiFe_sim} is truncated at
[Si/Fe]$=-0.5$. Nevertheless, the few model stars below [Si/Fe]$\sim 0$
do have a small contribution from electron capture SNe.

It should be noted that although the electron capture SNe indeed produce
a low Si yield and the fraction of low-Si stars enriched by this type of SN
is consistent with observations, the overall predicted abundance pattern
\citep{wan09} provides quite a poor fit to that observed in HE~$1424-0241$.
The situation improves if a few per mille of ejecta of an Fe-core collapse SN
is added to the gas. However, the fit to the light elements Na, Mg, and Al is still
poor. It is beyond the scope of this study to discuss the abundance pattern and
possible origin of HE~$1424-0241$ in detail. The interested reader is directed
to \citet{coh07}.

\subsubsection{A note on trends and observed scatter}\label{Sect:sscatter}

As discussed in \ref{Sect:avp} and the begining of \ref{Sect:mixing},
our results present not only a slope in [Si/Fe] with metallicity but
also a small cosmic scatter.

Trends, as well as scatters, are affected by the star formation and mixing
time scale of the ISM. Homogeneous chemical evolution models assume
instantaneous mixing. In these models, trends may, in the most metal-poor
regime, arise from the progenitor mass dependence of the SN yields. A
given abundance ratio, e.g., [Si/Fe], evolves with time, or metallicity, because
the most massive, short lived, SNe have a different [Si/Fe] yield ratio from those
of less massive, longer lived, SNe.

Mixing is, however, not instantaneous. In order to relax the assumption of
unphysically short mixing time scales, and still retain the small star-to-star
scatter observed in a number of abundance ratios, \citet{arn05} speculated
that the cooling time scale of metal-poor gas may be long enough for the
ISM to mix before subsequent generations of stars are able to form.
However, since the star-forming gas, in this scenario, always has to be well
mixed, such a ``global mixing'' would have difficulties to explain
any trends with metallicity, like the one reported here, \citep[see
also, e.g.,][]{cay04}, unless such trends are a result of a
metallicity-dependency of the SN yields. In the case of Si, the
conclusion is ambiguous. \citet{nom06}, predict a trend in [Si/Fe]
with metallicity which goes in the right direction, although with a
shallower slope than what is observed, while \citet{chi04} predict
almost no trend, however, with a very shallow slope in the opposite
direction.

An alternative explanation to the small observed scatter, without
invoking an unphysically short mixing time scale, is suggested by
\citet{bla10}.  They present a new stochastic chemical evolution
model in which stars are formed in clusters, as is known to be the
case in present-day star formation. In this scenario, the mixing
initially only occurs on a local scale. However, as a result of stars
being grouped together in clusters, the ejecta of $\ge 1$ SNe are
mixed together within each cluster, i.e, if the clusters are massive
enough to contain SNe. This may produce enough mixing to explain the
observations of, e.g., [Mg/Fe], while the large scatter observed for a
number of neutron-capture elements, e.g., [Ba/Fe] \citep{bur00,fra07}, can
still be accounted for. This will be discussed in a forthcoming paper.

\citet{kar05a} found trends with metallicity for certain
abundance ratios, while the scatter stayed small at all
metallicities, and, in particular cases, even decreased towards lower
metallicity. These trends are an effect of the local enrichment in
which different regions are enriched by SNe of different masses.
A similar effect was noticed by \citet{rya96}. If
the metal-poor star-forming gas were {\it not} very well mixed,
trends like these are to be expected, depending on the SN yields.
Note that the very same SN mass dependence could, in homogeneous
models, generate a trend with a different, or even opposite slope to that
in a stochastic, inhomogeneous model. Finally, a change in the IMF, e.g.
from a top-heavy to a Salpeter-like IMF, may also, possibly, generate a
trend with a non-zero slope. Clearly, in order to fully unravel the origins
of the observed trends at low metallicities, a deeper understanding of the
interplay between the mixing and cooling processes in the ISM is
necessary \citep{kar11}. This knowledge must be incorporated
in the modelling of chemical evolution.


\begin{acknowledgements}

We thank Dr. J.R. Shi for useful suggestions and discussions on NLTE
corrections. This work is supported by the NSFC under grant 10821061,
by the National Basic Rsearch Program of China under grant 2007CB815103,
and the Global Networks program of the University of Heidelberg. T.K.
is funded by ARC FF grant 0776384 through the
University of Sydney. T.K. is grateful to the Beecroft Institute for
Particle Astrophysics and Cosmology for their hospitality. A.J.K. acknowledges
support through grants by the Swedish Research Council (VR) and the Swedish
National Space Board (SNSB).P.S.B is aRoyal Swedish Academy of Sciences Research
Fellow supported by a grant from the Knut and Alice Wallenberg Foundation.
P.S.B also acknowledges additional support from the Swedish Research Council.
A number of comments and suggestions by an anonymous referee
helped improving the paper.

\end{acknowledgements}


\bibliographystyle{aa}
\bibliography{ref}


\Online

\begin{appendix} 

\onllongtab{3}{
\begin{landscape}
\begin{longtable}{lcccrrrrrrrr}
\caption{Abundance results of carbon and silicon. The last column is
the average of [Si/Fe] from two \ion{Si}{I} lines. If only upper
limit can be got from one line, taking the value of the other
line represents the average value.}\\
\hline\hline
     &        &       &    &    & \multicolumn{2}{c}{$\log \epsilon(\rm Si)_{\rm NLTE}$} & \multicolumn{2}{c}{[Si/H]$_{\rm NLTE}$} & \multicolumn{2}{c}{[Si/Fe]$_{\rm NLTE}$} & \\
star & [Fe/H] & $\log \epsilon(\rm C)$ & [C/H] & [C/Fe] & 3905 & 4103 & 3905 & 4103 & 3905 & 4103 & $\overline{\rm [Si/Fe]_{NLTE}}$ \\

 \hline
\endfirsthead
\caption{continued.}\\
\hline\hline
     &        &       &      &  & \multicolumn{2}{c}{$\log \varepsilon(\rm Si)_{\rm NLTE}$} & \multicolumn{2}{c}{[Si/H]$_{\rm NLTE}$} & \multicolumn{2}{c}{[Si/Fe]$_{\rm NLTE}$} & \\
star & [Fe/H] & $\log \varepsilon(\rm C)$ & [C/H] & [C/Fe] & 3905 & 4103 & 3905 & 4103 & 3905 & 4103 & $\overline{\rm [Si/Fe]_{NLTE}}$ \\
\hline
\endhead
\hline
\endfoot
\smallskip
CS22175-007  & --2.81  & 5.80  &  --2.72$\pm$0.14 &   0.09$\pm$0.16  &    5.16 & $<$5.19 &    --2.39$\pm$0.13   &  $<$--2.36$\pm$0.15  &     0.42$\pm$0.14  &    $<$0.45$\pm$0.16  &      0.42$\pm$0.14 \\
CS22186-023  & --2.72  & 6.00  &  --2.52$\pm$0.10 &   0.20$\pm$0.12  &    5.26 &    5.17 &    --2.29$\pm$0.09   &     --2.38$\pm$0.11  &     0.43$\pm$0.10  &       0.34$\pm$0.13  &      0.39$\pm$0.12 \\
CS22186-025  & --2.87  & 5.35  &  --3.17$\pm$0.15 & --0.30$\pm$0.17  &    5.22 &    5.28 &    --2.33$\pm$0.14   &     --2.27$\pm$0.16  &     0.54$\pm$0.15  &       0.60$\pm$0.17  &      0.57$\pm$0.17 \\
CS22886-042  & --2.68  & 5.71  &  --2.81$\pm$0.11 & --0.13$\pm$0.13  &    5.46 &    5.22 &    --2.09$\pm$0.10   &     --2.33$\pm$0.12  &     0.59$\pm$0.11  &       0.35$\pm$0.13  &      0.47$\pm$0.13 \\
CS22892-052  & --2.95  & 6.35  &  --2.17$\pm$0.11 &   0.78$\pm$0.13  &    5.31 &    5.13 &    --2.24$\pm$0.10   &     --2.42$\pm$0.12  &     0.71$\pm$0.11  &       0.53$\pm$0.13  &      0.62$\pm$0.13 \\
CS22945-028  & --2.66  & 6.11  &  --2.41$\pm$0.13 &   0.25$\pm$0.15  &    5.38 &    5.36 &    --2.17$\pm$0.12   &     --2.19$\pm$0.14  &     0.49$\pm$0.13  &       0.47$\pm$0.15  &      0.48$\pm$0.15 \\
CS22957-013  & --2.64  & 5.90  &  --2.62$\pm$0.12 &   0.02$\pm$0.14  &    5.34 &    5.34 &    --2.21$\pm$0.11   &     --2.21$\pm$0.13  &     0.43$\pm$0.12  &       0.43$\pm$0.14  &      0.43$\pm$0.14 \\
CS22958-083  & --2.79  & 6.28  &  --2.24$\pm$0.13 &   0.55$\pm$0.15  &    5.44 &    5.25 &    --2.11$\pm$0.12   &     --2.30$\pm$0.14  &     0.68$\pm$0.13  &       0.49$\pm$0.15  &      0.58$\pm$0.15 \\
\end{longtable}
\end{landscape}
}
\end{appendix}
\end{document}